\documentclass[12pt]{article}

\newlength{\dinwidth}
\newlength{\dinmargin}
\DeclareMathAlphabet{\mathsfsl}{OT1}{cmss}{m}{sl}
\DeclareMathAlphabet{\mathscr}{U}{eus}{m}{n}
\DeclareMathAlphabet{\matheur}{U}{eur}{m}{n}
\SetMathAlphabet{\matheur}{bold}{U}{eur}{b}{n}
\newcommand{\nn}{\nonumber \\}
\newcommand{\beq}{\begin{equation}} 
\newcommand{\eeq}{\end{equation}}
\newcommand{\bea}{\begin{eqnarray}}
\newcommand{\eea}{\end{eqnarray}}
\newcommand{\bruch}[2]{{\textstyle \frac{#1}{#2}}}
\newcommand{\ve}{\varepsilon}
\newcommand{\vf}{\varphi}
\newcommand{\ti}{\tau^\infty}

\newcommand{\A}{\alpha}
\newcommand{\B}{\beta}
\newcommand{\C}{t}
\newcommand{\GA}{\Gamma}

\newcommand{\M}{\mu}

\newcommand{\DC}{{\mathcal{D}}}

\newcommand{\VC}{{\mathcal{V}}}
\newcommand{\vh}{{\widehat{\mathcal{V}}}}

\newcommand{\TC}{{\mathcal{T}}}
\newcommand{\tTC}{{\widetilde{\mathcal{T}}}}
\newcommand{\FC}{{\mathcal{F}}}
\newcommand{\tFC}{{\widetilde {\mathcal{F}}}}
\newcommand{\MC}{{\mathcal{M}}}
\newcommand{\PC}{{\mathcal{P}}}
\newcommand{\QC}{{\mathcal{Q}}}
\newcommand{\tMC}{{\widetilde{\mathcal{M}}}}

\newcommand{\h}{ \frac{1}{2}}
\newcommand{\wh}{ \frac{1}{\sqrt{2}}}

\newcommand{\LWp}{{\rm SO}(1,1)_+}
\newcommand{\LWm}{{\rm SO}(1,1)_-}
\newcommand{\s}{{\rm SO(16)}}
\newcommand{\so}{{\rm SO}(1,2) \times {\rm SO}(16)}
\newcommand{\sa}{{\rm SO}(1,2) \times {\rm SO}(8)}
\newcommand{\sx}{{\rm SO}(1,1) \times {\rm SO}(16)^\infty}
\newcommand{\sxx}{{\rm SO}(16)^\infty}
\newcommand{\iso}{{\rm ISO}(9)}
\newcommand{\isx}{{\rm ISO}(16)^\infty}
\newcommand{\seps}{{\rm SO}(16)^\infty_{\varepsilon}}
\newcommand{\sy}{{\rm SO}(16)^{\infty \infty}}
\newcommand{\su}{{\rm SO}(1,3) \times {\rm SU}(8)}
\newcommand{\AmS}{{\protect\the\textfont2
  A\kern-.1667em\lower.5ex\hbox{M}\kern-.125emS}}

\begin {document}
\thispagestyle{empty}

% HIER EINFUEGEN: PREPRINTNUMMERN:
\begin{flushright} hep-th/9801090
                   \\  AEI--054
\end{flushright}
\vspace*{1.0cm}
\begin{center}
 {\LARGE \bf On M--Theory\footnotemark
% HIER EINFUEGEN: FUSSNOTENTEXT:
\footnotetext{Invited talks given at the 31st Ahrenshoop International 
Symposium on the Theory of Elementary Particles, Buckow, Germany,
2--6 September 1997, and the Discussion Meeting on Black Holes,
Bangalore, India, 8--10 December, 1997.}
}\\
 \vspace*{1cm}
 {{\bf Hermann Nicolai} \\
 \vspace*{6mm}
     Max-Planck-Institut f\"ur Gravitationsphysik\\
     Albert-Einstein-Institut \\
     Schlaatzweg 1, D-14473 Potsdam, Germany} \\
 \vspace*{1cm}
\begin{minipage}{11cm}\footnotesize
\textbf{Abstract:}
This contribution gives a personal view on recent attempts 
to find a unified framework for non-perturbative string theories, with 
special emphasis on the hidden symmetries of supergravity and their 
possible role in this endeavor. A reformulation of $d=11$ supergravity 
with enlarged tangent space symmetry $\so$ is discussed from this 
perspective, as well as an ansatz to construct yet further versions 
with $\sx$ and possibly even $\LWp\times\isx$ tangent space symmetry. 
It is suggested that upon ``third quantization'', dimensionally reduced 
maximal supergravity may have an equally important role to play 
in this unification as the dimensionally reduced maximally 
supersymmetric $SU(\infty)$ Yang Mills theory. 

\end{minipage}
\end{center}
\setcounter{footnote}{0}
\vspace*{0.2cm}
       
\section{Introduction}
Many theorists now believe that there is a unified framework for 
all string theories, which also accomodates $d=11$ 
supergravity \cite{CJS}. Much of the evidence for this elusive
theory, called ``M-Theory'' \cite{M}, is based on recent work 
on duality symmetries in string theory which suggests that all 
string theories are connected through a web of non-perturbative 
dualities \cite{duality}. Although it is unknown what M-theory really 
is, we can probably assert with some confidence $(i)$ that it will be 
a pregeometrical theory, in which space-time as we know it
will emerge as a secondary concept (which also means that
it makes little sense to claim that the theory ``lives'' in either
ten or eleven dimensions), and $(ii)$ that it should possess
a huge symmetry involving new and unexplored types of
Lie algebras (such as hyperbolic Kac Moody algebras), and perhaps 
other exotic structures such as quantum groups. In particular, 
the theory should be background independent and should be logically
deducible from a vast generalization of the principles underlying
general relativity.

According to a widely acclaimed recent proposal \cite{BFSS} M-Theory 
``is'' the $N\rightarrow\infty$ limit of the maximally supersymmetric 
quantum mechanical $SU(N)$ matrix model \cite{SUSY} (see \cite{dW} for 
recent reviews, points of view and comprehensive lists of references). 
This model had already appeared in an earlier investigation of 
the $d=11$ supermembrane \cite{BST} in a flat background in the light 
cone gauge \cite{dWHN}. Crucial steps in the developments leading 
up to this proposal were the discovery of Dirichlet $p$-branes and their
role in the description of non-perturbative string states \cite{Pol}, 
and the realization that the dynamics of an ensemble of such 
objects is described by dimensionally reduced supersymmetric 
Yang Mills theories \cite{Witten2}. Although there are a host of
unsolved problems in matrix theory, two central ones can perhaps
be singled out: one is the question whether the matrix model 
admits massless normalizable states for any $N$ (see \cite{massless} 
for recent work in this direction); the other is related to the 
still unproven existence of the $N\rightarrow\infty$ limit. This 
would have to be a weak limit in the sense of quantum field theory,
requiring the existence of a universal function $g=g(N)$ (the coupling 
constant of the $SU(N)$ matrix model) such that the limit 
$N\rightarrow\infty$ exists for all correlators. The existence of 
this limit would be equivalent to the renormalizability of the 
supermembrane \cite{dWHN}. However, even if these problems can be solved
eventually, important questions remain with regard to the assertions made
above: while matrix theory is pregeometrical in the sense that the target 
space coordinates are replaced by matrices, thus implying a kind of 
non-commutative geometry, the hidden exceptional symmetries of dimensionally 
reduced supergravities discovered long ago \cite{CJ,Julia1} are hard 
to come by (see \cite{EGKR} and references therein).
  
In the first part of this contribution, I will report on work 
\cite{NM}, which was motivated by recent advances in string theory as 
well as the possible existence of an Ashtekar-type canonical formulation 
of $d=11$ supergravity. Although at first sight our results, 
which build on earlier work of \cite{dewnic1,nic1}, may seem 
to be of little import for the issues raised above, I will argue 
that they could actually be relevant, assuming (as we do) 
that the success of the search for M-Theory will crucially 
depend on the identification of its underlying symmetries, 
and that the hidden exceptional symmetries of maximal supergravity 
theories may provide important clues as to where we should be looking. 
Namely, as shown in \cite{dewnic1,nic1}, the local symmetries of the 
dimensionally reduced theories can be partially ``lifted'' to 
eleven dimensions, indicating that these symmetries may have 
a role to play also in a wider context than that of dimensionally
reduced supergravity. The existence of alternative versions 
of $d=11$ supergravity, which, though equivalent on-shell to the
original version of \cite{CJS}, differ from it off-shell, 
suggests the existence of a novel kind of ``exceptional geometry'' 
for $d=11$ supergravity and the bigger theory containing it. 
This new geometry would be intimately tied to the special properties 
of the exceptional groups, and would be characterized by relations 
such as (\ref{id1})--(\ref{id4}) below, which have no analog in 
ordinary Riemannian geometry. The hope is, of course, that one may 
in this way gain valuable insights into what the (surely 
exceptional) geometry of M-Theory might look like, and that our
construction may provide a simplified model for it. After all, 
we do not even know what the basic physical concepts and
mathematical ``objects'' (matrices, BRST string functionals, spin
networks,...?) of such a theory should be, especially if it is to 
be a truly pregeometrical theory of quantum gravity.

The second part of this paper discusses the infinite dimensional
symmetries of $d=2$ supergravities \cite{Julia1,Julia2,BM,BMG,N,JN1,BJ}
and an ansatz that would incorporate them into the construction 
of \cite{NM,dewnic1,nic1}. The point of view adopted here is that the 
fundamental object of M-Theory could well be a kind of ``Unendlichbein'' 
belonging to an infinite dimensional coset space (cf. (\ref{coset3}) 
below), which would generalize the space $GL(4,{\bf R})/{\rm SO}(1,3)$
of general relativity. This bein would be acted upon from the 
right side by a huge extension of the Lorentz group, containing not 
only space-time, but also internal symmetries, and perhaps even 
local supersymmetries. For the left action, one would have to appeal 
to some kind of generalized covariance principle. An intriguing, 
but also puzzling, feature of the alternative formulations of 
$d=11$ supergravity is the apparent loss of manifest general 
covariance, as well as the precise significance of the global $E_{11-d}$ 
symmetries of the dimensionally reduced theories. This could mean 
that in the final formulation, general covariance will have to 
be replaced by something else.

The approach taken here is thus different from and arguably even
more speculative than current ideas based on matrix theory,
exploiting the observation that instead of dimensionally 
reducing the maximally extended {\em rigidly} supersymmetric 
theory to one dimension, one might equally well contemplate 
reducing the maximally extended {\em locally} supersymmetric 
theory to one (light-like $\equiv$ null) dimension. While matrix 
theory acquires an infinite number of degrees of freedom only in 
the $N\rightarrow\infty$ limit, the chirally reduced supergravity 
would have an infinite number from the outset, being one half of a 
field theory in two dimensions. The basic idea  
is then that upon quantization the latter might 
undergo a similarly far-reaching metamorphosis as the quantum 
mechanical matrix model, its physical states being
transmuted into ``target space'' degrees of freedom as in 
string theory \cite{nic2}. This proposal would amount to a third 
quantization of maximal ($N=16$) supergravity in two dimensions, 
where by ``third quantization'' I mean that the quantum treatment 
should take into account the gravitational degrees of freedom on 
the worldsheet, i.e. its (super)moduli for arbitrary genus. 
The model can be viewed as a very special example of $d=2$ quantum 
cosmology; with the appropriate vertex operator insertions
the resulting multiply connected $d=2$ ``universes''
can be alternatively interpreted as multistring scattering 
diagrams \cite{Mandel}. One attractive feature of this proposal 
is that it might naturally bring in $E_{10}$ as a kind of 
non-perturbative spectrum generating (rigid) symmetry acting on the 
third quantized Hilbert space, which would mix the worldsheet moduli 
with the propagating degrees of freedom. A drawback is that these 
theories are even harder to quantize than the matrix model
(see, however, \cite{KNS} and references therein).

\section{$\so$ invariant supergravity in eleven dimensions}
In \cite{dewnic1,nic1}, new versions of $d=11$ supergravity 
\cite{CJS} with local $\su$ and $\so$ tangent space symmetries, 
respectively, have been constructed. \cite{NM} develops these 
results further (for the $\so$ invariant version of \cite{nic1}),
and also discusses a hamiltonian formulation in terms of the 
new variables. In both versions the supersymmetry variations 
acquire a polynomial form from which the corresponding formulas 
for the maximal supergravities in four and three dimensions can 
be read off directly and without the need for complicated duality 
redefinitions. This reformulation can thus be regarded as a step 
towards the complete fusion of the bosonic degrees of freedom of $d=11$ 
supergravity (i.e. the elfbein $E_M^{~A}$ and the antisymmetric 
tensor $A_{MNP}$) in a way which is in harmony with the hidden 
symmetries of the dimensionally reduced theories.

For lack of space, and to exhibit the salient features as
clearly as possible I will restrict the discussion to the
bosonic sector. To derive the $\so$ invariant version of \cite{nic1,NM}
from the original formulation of $d=11$ supergravity, 
one first breaks the original tangent space symmetry SO(1,10) to 
its subgroup $\sa$  through a partial choice of gauge for the 
elfbein, and subsequently enlarges it again to $\so$ by introducing 
new gauge degrees of freedom. The symmetry enhancement 
of the transverse (helicity) group SO(9)$\, \subset \,$ SO(1,10)
to $\s$ requires suitable redefinitions of the bosonic and 
fermionic fields, or, more succinctly, their combination into 
tensors w.r.t. the new tangent space symmetry. The construction 
thus requires a 3+8 split of the $d=11$ coordinates and indices, 
implying a similar split for all tensors of the theory.
It is important, however, that the dependence on all eleven 
coordinates is retained throughout.

The elfbein and the three-index photon are thus combined into 
new objects covariant w.r.t. to the new tangent space symmetry. 
In the special Lorentz gauge preserving $\sa$ the elfbein
takes the form
\bea
E_M^{~A} = \left(\begin{array}{cc} 
            \Delta^{-1}e_\mu^{~a} & B_\mu^{~m} e_m^{~a}\\
            0& e_m^{~a}   \end{array} \right)
\label{11bein}
\eea
where curved $d=11$ indices are decomposed as $M=(\mu ,m)$ with 
$\M =0,1,2$ and $m= 3,...,10$ (with a similar decomposition of the flat 
indices), and $\Delta := {\rm det} \, e_m^{~a}$. In this gauge, 
the elfbein contains the (Weyl rescaled) dreibein and the Kaluza Klein 
vector $B_\mu{}^{m}$ both of which will be kept in the new formulation.
By contrast, the internal achtbein is replaced by a rectangular 248-bein 
$(e^{m}_{IJ},e^{m}_{A})$ containing the remaining ``matter-like'' 
degrees of freedom, where $([IJ],A)$ label the 248-dimensional 
adjoint representation of $E_8$ in the SO(16) decomposition.
This 248-bein, which in the reduction to three dimensions contains 
all the propagating bosonic matter degrees of freedom of $d=3,N=16$ 
supergravity, is defined in a special SO(16) gauge by 
\bea
(e^m_{IJ},e^m_A ) := \left\{ \begin{array}{ll}
     \Delta^{-1} e_a^{~m} \Gamma^a_{\A \dot \B} 
            & \mbox{if $[IJ]$ or $A = (\A \dot \B)$}\\
       0 & \mbox{otherwise}
       \end{array} \right.
\eea
where the $\s$ indices $IJ$ or $A$ are decomposed w.r.t. the diagonal 
subgroup ${\rm SO}(8)\equiv ({\rm SO}(8)\times {\rm SO}(8))_{diag}$  
of $\s$ (see \cite{nic1} for details). Being the inverse densitized 
internal achtbein contracted with an SO(8) $\Gamma$-matrix, this 
object is very much analogous to the inverse densitized triad in the 
framework of Ashtekar's reformulation of Einstein's theory \cite{A}.
Note that, due to its rectangularity, there does not exist an 
inverse for the 248-bein (nor is one needed for the supersymmetry
variations and the equations of motion!). In addition  we need 
the composite fields $(Q_{\M}^{IJ}, P_{\M}^{A})$ and 
$(Q_{m}^{IJ}, P_{m}^{A})$, which together make up an 
$E_8$ connection in eleven dimensions and whose explicit expressions 
in terms of the $d=11$ coefficients of anholonomity and the 
four-index field strength $F_{MNPQ}$ can be found in \cite{nic1}. 

The new geometry is encoded into algebraic constraints between 
the vielbein components, which are without analog in ordinary 
Riemannian geometry because they rely in an essential way on 
special properties of the exceptional group $E_8$. We have
\bea
e^m_A e^n_A - \bruch{1}{2}e^m_{IJ} e^n_{IJ} = 0 \label{id1}
\eea
and
\bea
\GA^{IJ}_{AB} \Big( e^m_B e^n_{IJ} - e^n_B e^m_{IJ} \Big) = 0  \qquad
\GA^{IJ}_{AB} e^m_A e^n_B + 4 e^m_{K[I} e^n_{J]K} = 0 \label{id2}
\eea
where $\Gamma^I_{A\dot A}$ are the standard SO(16) $\GA$-matrices and 
$\GA_{AB}^{IJ}\equiv (\GA^{[I} \GA^{J]})_{AB}$, etc.; the minus 
sign in (\ref{id1}) reflects the fact that we are dealing with
the maximally non-compact form $E_{8(+8)}$. While the SO(16) covariance 
of these equations is manifest, it turns out, remarkably, that 
they are also covariant under $E_8$. Obviously, (\ref{id1}) 
and (\ref{id2}) correspond to the singlet and the adjoint 
representations of $E_8$. More complicated are the following
relations transforming in the $\bf 3875$ representation of $E_8$ 
\bea
e^{(m}_{IK} e^{n)}_{JK} - \bruch{1}{16} \delta_{IJ} 
e^m_{KL} e^n_{KL}  &=& 0   \nn
\GA^K_{\dot A B} e^{(m}_B e^{n)}_{IK} - \bruch{1}{14}
\GA^{IKL}_{\dot A B} e^{(m}_B e^{n)}_{KL} &=& 0 \nn
e^{(m}_{[IJ} e^{n)}_{KL]} + \bruch{1}{24}
e^m_A \GA^{IJKL}_{AB} e^n_B  &=& 0 \label{id4}
\eea
Yet another set of relations involves the $\bf 27000$ representation 
of $E_8$ \cite{NM}.

The 248-bein and the new connection fields are subject to 
a ``vielbein postulate" similar to the usual vielbein postulate 
stating the covariant constancy of the vielbein w.r.t. 
to generally covariant and Lorentz covariant derivative: 
\bea
(\partial_\mu - B_\mu^{~n}\partial_n) e_{IJ}^{m} +
\partial_{n} B_{\mu}{}^{n} e^{m}_{IJ} +
\partial_{n}B_{\mu}{}^{m} e^{n}_{IJ} + 2\, {{Q_\mu}^K}_{[I} e^m_{J]K}
 + P_\mu^A \Gamma^{IJ}_{AB} e_m^B  &=& 0 \nn
(\partial_\mu - B_\mu^{~n} \partial_n) e_{A}^{m} +
\partial_{n} B_{\mu}{}^{m} e^{n}_{A} +
\partial_{n}B_{\mu}{}^{n} e^{m}_{A}
 + \bruch{1}{4} Q_\mu^{IJ} \GA^{IJ}_{AB} e^m_B -
\bruch{1}{2} \Gamma^{IJ}_{AB} P_{\mu}^{B} e^{m}_{IJ} &=& 0 \nn
\partial_m e^{n}_{IJ} + 2\, {{Q_m}^K}_{[I} e^n_{J]K} 
 +P_{m}^{A} \Gamma^{IJ}_{AB} e^n_B & = & 0 \nn
\partial_m e^{n}_{A} + \bruch{1}{4}Q_m^{IJ} \GA^{IJ}_{AB} e^n_B
-\bruch{1}{2} \Gamma^{IJ}_{AB} P_{m}^{B} e^{n}_{IJ} & = & 0
\label{VVP4}
\eea
Like (\ref{id1})--(\ref{id4}), these relations are $E_8$ covariant. It 
must be stressed, however, that the full theory of course does not respect
$E_8$ invariance. A puzzling feature of (\ref{VVP4}) is that the 
covariantization w.r.t. an affine connection is ``missing'' in these 
equations, even though the theory is still invariant under $d=11$ 
coordinate transformations. One can now show that the supersymmetry 
variations of $d=11$ supergravity can be entirely expressed in 
terms of these new variables (and their fermionic partners). 

The reduction of $d=11$ supergravity to three dimensions yields 
$d=3, N=16$ supergravity \cite{MS}, and is accomplished rather easily, 
since no duality redefinitions are needed any more, unlike in \cite{CJ}. 
The propagating bosonic degrees of freedom in three dimensions are 
all scalar, and combine into a matrix $\VC(x)$, which is an element 
of a non-compact $E_{8(+8)}/\s$ coset space, and whose dynamics 
is governed by a non-linear $\sigma$-model coupled to 
$d=3$ gravity. The identification of the 248-bein with the 
$\sigma$-model field $\VC\in E_8$ is given by 
\bea 
e^m_{IJ} = \bruch{1}{60}{\rm Tr} \, \big( Z^m \VC X^{IJ} \VC^{-1} \big)
\qquad
e^m_A = \bruch{1}{60}{\rm Tr} \, \big( Z^m \VC Y^A \VC^{-1} \big)
\label{bein1}
\eea
where $X^{IJ}$ and $Y^A$ are the compact and non-compact 
generators of $E_8$, respectively, and where the $Z^m$ for 
$m=3,...,10$ are eight non-compact commuting generators obeying 
${\rm Tr} (Z^m Z^n) = 0$ for all $m$ and $n$ (the existence of 
eight such generators is a consequence of the fact that the coset 
space $E_{8(+8)}/\s$ has real rank 8 and therefore admits an 
eight-dimensional maximal flat and totally geodesic submanifold \cite{H}). 
This reduction provides a ``model'' for the exceptional geometry, where 
the relations (\ref{id1})--(\ref{VVP4}) can be tested by means of 
completeness relations for the $E_8$ Lie algebra generators in 
the adjoint representation. Of course, this is not much of a 
test since all dependence on the internal coordinates is 
dropped in (\ref{bein1}), and the terms involving  $B_\mu^{~m}$ 
disappear altogether. It would be desirable to find other ``models" 
with non-trivial dependence on the internal coordinates. The only 
example of this type so far is provided by the $S^7$ truncation  
of $d=11$ supergravity for the $\su$ invariant version of $d=11$ 
supergravity \cite{dewnic2}.

\section{More Symmetries}
The emergence of hidden symmetries of the exceptional type in 
extended supergravities \cite{CJ} was a remarkable and, at the 
time, quite unexpected discovery. It took some effort to show
that the general pattern continues when one descends to $d=2$ 
and that the hidden symmetries become infinite dimensional 
\cite{Julia1,Julia2,BM,BMG,N,JN1,BJ}, generalizing the Geroch 
group of general relativity \cite{Geroch}. As we will see, even the 
coset structure remains, although the mathematical objects one 
deals with become a lot more delicate. The fact that the 
construction described above works with a 4+7 and 3+8 split 
of the indices suggests that we should be able to go even further 
and to construct versions of $d=11$ supergravity with infinite 
dimensional tangent space symmetries, which would be based on a
2+9 or even a 1+10 split of the indices. This would also be desirable
in view of the fact that the new versions are ``simple'' only in 
their internal sectors. The general strategy is thus to further 
enlarge the internal sector by absorbing more and more degrees of 
freedom into it, such that in the final step corresponding to a 
1+10 split, only an einbein is left in the low dimensional sector. 
Although the actual elaboration of these ideas has to be left to 
future work, I will try to give at least a flavor of some anticipated 
key features.

\subsection{Reduction to two dimensions}
Let us first recall some facts about dimensional reduction 
of maximal supergravity to two dimensions. Following the 
empirical rules of dimensional reduction one is led to predict 
$E_9 = E_8^{(1)}$ as a symmetry for the dimensional reduction 
of $d=11$ supergravity to two dimensions \cite{Julia1}.
This expectation is borne out by the existence of a linear 
system for maximal $N=16$ supergravity in two dimensions 
\cite{nic2,NW} (see \cite{BelZak,BM} for the bosonic theory). 
The linear system requires the introduction of an extra
``spectral" parameter $\C$, and the extension of the $\sigma$-model 
matrix $\VC (x)$ to a matrix $\vh(x;\C)$ depending on this extra 
parameter $\C$, as is generally the case for integrable systems
in two dimensions. An unusual feature is that, due to the
presence of gravitational degrees of freedom, this parameter becomes 
coordinate dependent, i.e. we have $\C=\C(x;w)$, where $w$ is an 
integration constant, sometimes referred to as the ``constant 
spectral parameter'' whereas $\C$ itself is called the ``variable 
spectral parameter''.

Here, we are mainly concerned with the symmetry aspects of this 
system, and with what they can teach us about the $d=11$ theory 
itself. The coset structure of the higher dimensional theories has 
a natural continuation in two dimensions, with the only difference
that the symmetry groups are infinite dimensional. This property 
is manifest from the transformation properties of the linear system 
matrix $\vh$, with a global affine symmetry acting from the left,
and a local symmetry corresponding to some ``maximal compact'' 
subgroup acting from the right:
\bea
\vh (x;\C) \longrightarrow g(w) \vh(x;\C) h(x;\C)
\eea
Here $g(w)\in E_9$ with affine parameter $w$, and the 
subgroup to which $h(x;\C)$ belongs is characterized as follows 
\cite{Julia2,BM}. Let $\tau$ be the involution characterizing the 
coset space $E_{8(+8)}/\s$: then $h(\C)\in\seps$ is defined to 
consist of all $\ti$ invariant elements of $E_9$, where the 
extended involution $\ti$ is defined by 
$\ti(h(\C)):= \tau h(\ve \C^{-1})$, with $\ve=+1$ (or $-1$) 
for a Lorentzian (Euclidean) worldsheet. For $\ve=1$, which is 
the case we are mainly interested in, we will write 
$\sxx \equiv \seps$. We also note that $\seps$ is different
from the affine extension of $\s$ for either choice of sign.

What has been achieved by the coset space description is
the following: by representing the ``moduli space of solutions'' 
$\MC$ (of the bosonic equations of motion of $d=11$ supergravity 
with nine commuting space-like Killing vectors) as
\bea
\MC = \frac{{\rm solutions \, of \, field \, equations}}%
             {{\rm diffeomorphisms}} 
    = \frac{E_9}{\sxx} \label{coset1}
\eea
we have managed to endow this space, which a priori is very 
complicated, with a group theoretic structure, that makes it 
much easier to handle. In particular, the integrability of the 
system is directly linked to the fact that $\MC$ possesses
an infinite dimensional ``isometry group'' $E_9$. The introduction
of infinitely many gauge degrees of freedom embodied in the
subgroup $\sxx$ linearizes and localizes the action of this
isometry group on the space of solutions. Of course, 
in making such statements, one should keep in mind that a
mathematically rigorous construction of such spaces is a 
thorny problem. This is likewise true for the infinite dimensional
groups\footnote{For instance, the Geroch group can be defined
rigorously to consist of all maps from the complex $w$ plane 
to $SL(2,{\bf R})$ with meromorphic entries. With this definition,
one obtains all multisoliton solutions of Einstein's equations, 
and on this solution space the group acts transitively by construction. 
Whether this is the right choice or not is then a matter of 
physics, not mathematics.} and their associated Lie algebras; 
the latter being infinite dimensional vector spaces, there are 
myriad ways of equiping them with a topology. We here take the 
liberty of ignoring these subleties, not least because 
these spaces ultimately will have to be ``quantized'' anyway.

There is a second way of defining the Lie algebra 
of $\seps$ which relies on the Chevalley-Serre presentation. Given a 
finite dimensional non-compact Lie group $G$ with maximal compact 
subgroup $H$, a necessary condition for this prescription to work 
is that ${\rm dim} \,H = \frac12 ({\rm dim}\, G - {\rm rank}\, G)$,
and we will subsequently extend this prescription to the infinite
Lie group. Let us first recall that any (finite or infinite 
dimensional) Kac Moody algebra is recursively defined in terms 
of multiple commutators of the Chevalley generators subject to 
certain relations \cite{Kac}. More specifically, given a 
Cartan matrix $A_{ij}$ and the associated Dynkin diagram,
one starts from a set of $sl(2,{\bf R})$ generators $\{e_i,f_i,h_i\}$, 
one for each node of the Dynkin diagram, which in addition to
the standard $sl(2,{\bf R})$ commutation relations 
\bea
[h_i , h_j] = 0 \qquad  [e_i , f_j] = \delta_{ij} h_j \nonumber
\eea
\bea
[h_i , e_j] = A_{ij} e_j \qquad  [h_i , f_j ] = -A_{ij} f_j \label{Serre1}
\eea
are subject to the multilinear Serre relations
\bea
[e_i,[e_i,...[e_i,e_j]...]]  = 0  \qquad
[f_i,[f_i,...[f_i,f_j]...]]  = 0 \label{Serre2}
\eea
where the commutators are $(1-A_{ij})$-fold ones. The Lie algebra 
is then by definition the linear span of all multiple commutators 
which do not vanish by virtue of these relations.

To define the subalgebra $\seps$, we first recall that the
Chevalley involution $\theta$ is defined by
\bea
\theta(e_i) = -f_i \qquad
\theta(f_i) = -e_i \qquad
\theta(h_i) = -h_i
\eea
This involution, like the ones to be introduced below, leaves 
invariant the defining relations (\ref{Serre1}) and (\ref{Serre2}) 
of the Kac Moody algebra, and extends to the whole Lie algebra 
via the formula $\theta ([x,y])=[\theta (x),\theta (y)]$.
It is not difficult to see that, for $E_8$ (and also 
for $sl(n,{\bf R})$), we have $\tau = \theta$, and the maximal compact 
subalgebras defined above correspond to the subalgebras generated by 
the multiple commutators of the $\theta$ invariant elements $(e_i-f_i)$ 
in both cases. The trick is now to carry over this definition to 
the affine extension, whose associated Cartan matrix has a zero 
eigenvalue. To do this, however, we need a slight generalization
of the above definition; for this purpose, we consider involutions
$\omega$ that can be represented as products of the form
\bea
\omega = \theta \cdot s \label{involution}
\eea
where the involution $s$ acts as
\bea
s (e_i) = s_i e_i \qquad
s (f_i) = s_i f_i \qquad
s (h_i) = h_i    \label{s}
\eea
with $s_i = \pm 1$. It is important that different choices of $s_i$ do 
not necessarily lead to inequivalent involutions (the general problem 
of classifying the involutive automorphisms of infinite dimensional 
Kac Moody algebras has so far not been completely solved, see e.g. 
\cite{inv}\footnote{I am very grateful to C.~Daboul for helpful
discussions on this topic.}). In particular for $E_9$, 
which is obtained from $E_8$ by adjoining another set $\{e_0,f_0,h_0\}$ 
of Chevalley generators, we take $s_i =1$ for all $i\geq 1$, whereas 
$s_0 = \ve$, with $\ve$ as before, i.e. $\ve=+1$ (or $-1$) for Lorentzian 
(Euclidean) worldsheet. Thus, on the extended Chevalley generators,
\bea
\omega (e_0) = - \ve f_0 \qquad
\omega (f_0) = - \ve e_0 \qquad
\omega (h_0) = -h_0        \label{involution1}
\eea
With this choice, the involution $\omega$ coincides with the
involutions defined before for the respective choices of $\ve$,
i.e. $\omega = \ti$, and therefore the invariant subgroups are 
the same, too. For $\ve = 1$, the involution $\omega$ defines an 
infinite dimensional ``maximal compact'' subalgebra consisting of 
all the negative norm elements w.r.t. to the standard bilinear form 
\bea 
\langle e_i| f_j\rangle = \delta_{ij} \qquad
\langle h_i| h_j\rangle = A_{ij} 
\eea
(the norm of any given multiple commutator can be
determined recursively from the fundamental relation 
$\langle[x,y]|z\rangle = \langle x|[y,z]\rangle$). The notion of 
``compactness'' here is thus algebraic, not topological: the subgroup 
$\sxx$ will not be compact in the topological sense (recall the 
well known example of the unit ball in an infinite dimensional 
Hilbert space, which is bounded but not compact in the norm topology). 
On the other hand, for $\ve=-1$, the group $\seps$ is not even compact 
in the algebraic sense, as $e_0+f_0$ has positive norm. However,
this is in accord with the expectation that $\seps$ should contain
the (non-compact) group SO(1,8) rather than SO(9) if one of the
compactified dimensions is time-like.

\subsection{$2+9$ split}
Let us now consider the extension of the results described in
section 2 to the situation corresponding to a 2+9 split of the 
indices. Elevating the local symmetries of $N=16$ supergravity  
from two to eleven dimensions would require the existence 
of yet another extension of the theory, for which the Lorentz 
group SO(1,10) is replaced by $\sx$; the subgroup $\sxx$ can 
be interpreted as an extension of the transverse group SO(9) 
in eleven dimensions. Taking the hints from (\ref{11bein}),
we would now decompose the elfbein into a zweibein and nine 
Kaluza Klein vectors $B_\mu^{~m}$ (with $m=2,...,10$).
The remaining internal neunbein would have to be replaced by 
an ``Unendlichbein'' $\big(e^m_{IJ}(x;\C),e^m_A(x;\C)\big)$, 
depending on a spectral parameter $\C$, necessary to 
parametrize the infinite dimensional extension of the symmetry 
group. However, in eleven dimensions, there is no anolog of the 
dualization mechanism, which would ensure that despite the 
existence of infinitely many dual potentials, there are only finitely 
many physical degrees of freedom. This indicates that if the 
construction works it will take us beyond $d=11$ supergravity. 

Some constraints on the geometry can be deduced from the
requirement that in the dimensional reduction to $d=2$, there 
should exist a formula analogous to (\ref{bein1}), but with 
$\VC$ replaced by the linear system matrix $\vh$, or 
possibly even the enlarged linear system of \cite{JN1}. Evidently,
we would need a ninth nilpotent generator to complement
the $Z^m$'s of (\ref{bein1}); an obvious candidate is the central
charge generator $c$, since it obeys $\langle c|c \rangle =
\langle c| Z^m \rangle = 0$ for all $m=3,...,10$. The parameter 
$\C$, introduced somewhat ad hoc for the parametrization of the 
unendlichbein, must obviously coincide with the spectral 
parameter of the $d=2$ theory, and the generalized
``unendlichbein postulate'' should evidently reduce to the 
linear system of $d=2$ supergravity in this reduction. To write 
it down, we need to generalize the connection coefficients
appearing in the linear system. The latter are given by
\bea
\QC_\mu^{IJ} = Q_\mu^{IJ} + \dots \qquad
\PC_\mu^A = \frac{1+\C^2}{1-\C^2} P_\mu^A +
   \frac{2\C}{1-\C^2} \ve_{\mu \nu} P^{\nu A} + \dots  \label{conn}
\eea
with $Q_\mu^{IJ}$ and $P_\mu^A$ as before; the dots indicate 
$\C$ dependent fermionic contributions which we omit. 
A very important difference with section 2, where the tangent space 
symmetry was still finite dimensional, is that the Lie algebra 
of $\sxx$ also involves the $P$'s, and not only the $Q$'s. More 
specifically, from the $\C$ dependence of the dimensionally reduced 
connections in (\ref{conn}) we infer that the connections 
$(\QC_M^{IJ}(x;\C), \PC_M^A(x;\C))$ constitute an $\sxx$ (and not
an $E_9$) gauge connection. This means that the covariantizations in the
generalized vielbein postulate are now in precise correpondence
with the local symmetries, in contrast with the relations
(\ref{VVP4}) which look $E_8$ covariant, whereas the full theory
is invariant only under local $\s$.

To write down an ansatz, we put
\bea
\DC_\mu := \partial_\mu - B_\mu^{~n} \partial_n + \dots
\eea
where the dots stand for terms involving derivatives of the Kaluza
Klein vector fields. Then the generalization of (\ref{VVP4}) 
should read
\bea
\DC_\mu e_{IJ}^{m}(\C)
 + 2\, {{\QC_\mu}^K}_{[I} (\C)e^m_{J]K}(\C) 
 + \, \PC_\mu^A(\C) \Gamma^{IJ}_{AB} e^m_B (\C)  &=& 0 \nn
\DC_\mu e_{A}^{m}(\C) +
 \bruch{1}{4} \QC_\mu^{IJ}(\C) \GA^{IJ}_{AB} e^m_B(\C) 
-\, \bruch{1}{2} \Gamma^{IJ}_{AB} \PC_{\mu}^{B}(\C) e^{m}_{IJ}(\C) &=& 0 \nn
\partial_m e^{n}_{IJ}(\C) + 2\, {{\QC_m}^K}_{[I}(\C) e^n_{J]K}(\C)
 +\, \PC_{m}^{A}(\C) \Gamma^{IJ}_{AB} e^n_B(\C) & = & 0 \nn
\partial_m e^{n}_{A}(\C) + \bruch{1}{4}\QC_m^{IJ}(\C) \GA^{IJ}_{AB} e^n_B(\C)
-\bruch{1}{2} \, \Gamma^{IJ}_{AB} \PC_{m}^{B}(\C) e^{n}_{IJ}(\C) & = & 0
\label{VVP5}
\eea
Of course, the challenge is now to find explicit expressions
for the internal components $\QC_m^{IJ}(x;\C)$ and $\PC_m^A(x;\C)$, 
such that (\ref{VVP5}) can be interpreted as a $d=11$ generalization 
of the linear system of dimensionally reduced supergravity.
Another obvious question concerns the fermionic partners of
the unendlichbein: in two dimensions, the linear system matrix
contains all degrees of freedom, including the fermionic ones,
and the local $N=16$ supersymmetry can be bosonized into a local
$\sxx$ gauge transformation \cite{NW}. Could this mean
that there is a kind of bosonization in eleven dimensions or
M-Theory? This idea may not be as outlandish as it sounds because
a truly pregeometrical theory might be subject to a kind of
``pre-statistics'', such that the distinction between bosons and 
fermions arises only through a process of spontaneous symmetry breaking.

\section{Yet more symmetries?}
In 1982, B.~Julia conjectured that the dimensional reduction
of maximal supergravity to one dimension should be invariant
under a further extension of the $E$-series, namely (a non-compact
form of) the hyperbolic Kac Moody algebra $E_{10}$ obtained 
by adjoining another set $\{e_{-1}, f_{-1}, h_{-1}\}$ 
of Chevalley generators to those of $E_9$
\cite{Julia3}\footnote{The existence of a maximal dimension
for supergravity \cite{Nahm} would thus be correlated with 
the existence of a ``maximally extended'' hyperbolic Kac Moody algebra, 
which might thus explain the occurrence of maximum spin 2 
for massless gauge particles in nature.}. As shown in \cite{nic5}, 
the last step of the reduction requires a null reduction if the 
affine symmetry of the $d=2$ theory is not to be lost. The reason 
is that the infinite dimensional affine symmetries of 
the $d=2$ theories always involve dualizations of the type
\bea
\partial_\mu \vf = 
\ve_{\mu \nu} \partial^\nu \tilde \vf 
\eea
(in actual fact, there are more scalar fields, and the duality 
relation becomes non-linear, which is why one ends up with infinitely
many dual potentials for each scalar degree of freedom).
Dimensional reduction w.r.t. to a Killing vector $\xi^\mu$ 
amounts to imposing the condition $\xi^\mu \partial_\mu \equiv 0$ 
on {\it all} fields, including dual potentials. Hence, 
\bea
\xi^\mu \partial_\mu \vf = 0  \quad , \quad
\xi^\mu \partial_\mu \tilde \vf \equiv \eta^\mu \partial_\mu \vf = 0
\eea
where $\eta^\mu \equiv \ve^{\mu \nu} \xi_\nu$. If $\xi^\mu$ and
$\eta^\mu$ are linearly independent, this constraint would force
all fields to be constant, which is clearly too strong a 
requirement. Hence we must demand that $\xi^\mu$ and $\eta^\mu$
are collinear, which implies 
\bea
 \xi^\mu \xi_\mu =0 , 
\eea
i.e. the Killing vector must be null. Starting from this observation, it 
was shown in \cite{nic5} that the Matzner Misner $sl(2,{\bf R})$ symmetry 
of pure gravity can be formally extended to an $sl(3,{\bf R})$ 
algebra in the reduction of the vierbein from four to one dimensions.
Combining this $sl(3,{\bf R})$ with the Ehlers $sl(2,{\bf R})$ 
of ordinary gravity, or with the $E_8$ symmetry of maximal 
supergravity in three dimensions, one is led to the hyperbolic 
algebra ${\FC}_3$ \cite{FF} for ordinary gravity, and 
to $E_{10}$ for maximal supergravity. The transformations 
realizing the action of the Chevalley generators on the vierbein
components can be worked out explicitly, and the Serre relations
can be formally verified \cite{nic5} (for $E_{10}$, this was 
shown more recently in \cite{Mizo}). 

There is thus some evidence for the emergence of hyperbolic 
Kac Moody algebras in the reduction to one null dimension, but the 
difficult open question that remains is what the configuration space 
is on which this huge symmetry acts. This space is expected to be much 
bigger than the coset space (\ref{coset1}). Now, already for the $d=2$
reduction there are extra degrees of freedom that must be taken into 
account in addition to the propagating degrees of freedom. Namely, the full 
moduli space involving all bosonic degrees of freedom should also 
include the moduli of the zweibein, which are not contained 
in (\ref{coset1}). For each point on the worldsheet, the zweibein 
is an element of the coset space ${\rm GL}(2,{\bf R})/{\rm SO}(1,1)$; 
although it has no local degrees of freedom any more, it still 
contains the global information about the conformal structure of 
the world sheet $\Sigma$. Consequently, we should consider 
the Teichm\"uller space
\bea
\TC = 
\frac{ \{e_\mu^{~\A}(x) \, | \, x\in \Sigma \} }%
{{\rm SO}(1,1)\times{\rm Weyl} (\Sigma)\times{\rm Diff}_0 (\Sigma)}
\label{coset2}
\eea
as part of the configuration space of the theory (see \cite{Verlinde}
for a detailed description of $\TC$). In fact, we should even allow 
for arbitrary genus of the worldsheet, and replace $\TC$ by
the ``universal Teichm\"uller space'' $\tTC$. This infinite
dimensional space can be viewed as the configuration space 
space of non-perturbative string theory \cite{FS}. For the
models under consideration here, however, even $\tTC$ is not 
big enough, as we must also take into account the dilaton $\rho$
and the non-propagating Kaluza Klein vector fields in two 
dimensions. For the former, a coset space description was proposed 
in \cite{JN1}. On the other hand, the Kaluza Klein vectors
and the cosmological constant they could generate in two dimensions 
have been largely ignored in the literature. Even if one sets 
their field strengths equal to zero (there are arguments that the 
Geroch group, and hence infinite duality symmetries, are incompatible 
with a nonzero cosmological constant in two dimensions), there still 
remain topological degrees of freedom for higher genus world sheets.
 
The existence of inequivalent conformal structures is evidently 
important for the null reductions, as the former are in one-to-one
correspondence with the latter. Put differently, the inequivalent 
null reductions are precisely parametrized by the space
(\ref{coset2}). The extended symmetries should thus not only
act on one special null reduction (set of plane wave solutions
of Einstein's equations), but relate different reductions. Indeed,
it was argued in \cite{Mizo} that, for a toroidal worldsheet,
the new $sl(2,{\bf R})$ transformations associated 
with the over-extended Chevalley generators change the conformal 
structure, but only for non-vanishing holonomies of the Kaluza 
Klein vector fields on the worldsheet. This indicates that the 
non-trivial realization of the hyperbolic symmetry requires
the consideration of non-trivial worldsheet topologies.
The dimensionally reduced theory thereby retains a memory of its
two-dimensional ancestor. It is therefore remarkable that,
at least for isomonodromic solutions of Einstein's theory,
the $d=2$ theory exhibits a factorization of the equations of motion
akin to, but more subtle than the holomorphic factorization of
conformal field theories \cite{KN}. In other words, there may be 
a way to think of the $d=2$ theory as being composed of two chiral 
halves just as for the closed string. Consequently, a truncation to one 
null dimension may not be necessary after all if the theory 
factorizes all by itself.

In summary, what we are after here is a group theoretic unification of all 
these moduli spaces that would be analogous to (\ref{coset1}) above, and 
fuse the matter and the topological degrees of freedom. No such description 
seems to be available for (\ref{coset2}) (or $\tTC$), and it is conceivable 
that only the total moduli space $\tMC$ containing both $\MC$ and $\tTC$ 
as well as the dilaton and the Kaluza Klein, and perhaps even the 
fermionic, degrees of freedom is amenable to such an interpretation.
Extrapolating the previous results, we are thus led to consider
coset spaces $E_{10}/H$ with $\sxx\subset H \subset E_{10}$.
As before, the introduction of the infinitely many spurious 
degrees of freedom associated with the gauge group $H$ would
be necessary in order to ``linearize'' the action of $E_{10}$.

What are the choices for $H$? One possibility would be to follow 
the procedure of the foregoing section, and to define 
$H = \sy \subset E_{10}$ in analogy with $\sxx\subset E_9$ by
taking its associated Lie algebra to be the linear span of all 
$\omega$ invariant combinations of $E_{10}$ Lie algebra elements. 
To extend the affine involution to the full hyperbolic algebra, 
we would again invoke (\ref{involution}), setting $\ve=+1$ in 
(\ref{involution1}) (since we now assume the worldsheet to be
Lorentzian), which leaves us with the two choices $s_{-1}=\pm 1$.
For $s_{-1}=+1$ we would get the ``maximal compact'' subalgebra of 
$E_{10}$, corresponding to the compactification of ten 
spacelike dimensions. A subtlety here is that a definition 
in terms of the standard bilinear form is no longer possible, 
unlike for affine and finite algebras, as this would now also 
include part of the Cartan subalgebra of $E_{10}$: due to the 
existence of a negative eigenvalue of the $E_{10}$ Cartan matrix, 
there exists a negative norm element $\sum _i n_i h_i$ of the Cartan 
subalgebra, which would have to be excluded from the definition of
$H$ (cf. the footnote on p.~438 of \cite{JN1}). The alternative choice 
$s_{-1}=-1$ would correspond to reduction on a 9+1 torus. 

However, for the null reduction advocated here, physical reasoning 
motivates us to propose yet another choice for $H$. Namely,
in this case, $H$ should contain the group 
$\iso \subset {\rm SO}(1,10)$ leaving 
invariant a null vector in eleven dimensions \cite{JN2}. To identify
the relevant parabolic subgroup of $E_{10}$, which we denote by
$\isx$, we recall \cite{nic5} that the over-extended Chevalley 
generators correspond to the matrices 
\bea
 e_{-1} = \wh \left( \begin{array}{clcr}
                0 & 0 & 1   \\
                0 & 0 & 1    \\
                0 & 0 & 0   \end{array}\right)   \qquad
 f_{-1} = \wh \left( \begin{array}{clcr}
                0 & 0 & 0 \\
                0 & 0 & 0 \\
                1 & 1 & 0 \end{array}\right)  \qquad
 h_{-1}  = \h \left( \begin{array}{clcr}
                 1 & 1 & 0 \\
                 1 & 1 & 0 \\
                 0 & 0 & -2 \end{array}\right)
\eea
in a notation where we only write out the components acting
on the $0,1,2$ components of the elfbein, with all other entries 
vanishing. Evidently, we have $h_{-1}=d-c_-$ with
\bea
 d  =  \left( \begin{array}{clcr}
                 0 & 0 & 0 \\
                 0 & 0 & 0 \\
                 0 & 0 & -1 \end{array}\right)  \qquad
 c_-  = - \h \left( \begin{array}{clcr}
                 1 & 1 & 0 \\
                 1 & 1 & 0 \\
                 0 & 0 & 0 \end{array}\right) 
\eea
where $d$ is the scaling operator on the dilaton $\rho$ , and
$c_-$ is the central charge, alias the ``level counting operator'' of 
$E_{10}$, obeying $[c_-,e_{-1}]= -e_{-1}$ and $[c_-,f_{-1}]= +f_{-1}$
(and having vanishing commutators with all other Chevalley generators).
Writing
\bea
c_\pm := - \h \left( \begin{array}{clcr}
                    1 & 0 & 0 \\
                    0 & 1 & 0 \\
                    0 & 0 & 0 \end{array}\right)
         \pm \h \left( \begin{array}{clcr}
                    0 & 1 & 0 \\
                    1 & 0 & 0 \\
                    0 & 0 & 0 \end{array}\right)
\eea
we see that the first matrix on the right scales the conformal
factor, generating Weyl transformations (called ${\rm Weyl}(\Sigma)$
in (\ref{coset2})) on the zweibein, while the second generates
the local SO(1,1) Lorentz transformations. In a lightcone basis,
these symmetries factorize on the zweibein, which decomposes into
two chiral einbeine. Consequently, Weyl transformations and local
SO(1,1) can be combined into two groups ${\rm SO}(1,1)_\pm$ with
respective generators $c_\pm$, and which act separately
on the chiral einbeine. One of these, $\LWm$ (generated by $c_-$),
becomes part of $E_{10}$. The other, $\LWp$, acts on the residual 
einbein and can be used to eliminate it by gauging it to one. 
Since $c_\pm$ acts in the same way on the conformal factor, we also 
recover the result of \cite{Julia2}.

We wish to include both $\iso$ and $\LWm$ into the enlarged local
symmetry $H=\isx$, and thereby unify the longitudinal symmetries 
with the ``transversal'' group $\sxx$ discussed before. Accordingly,
we define $\isx$ to be the algebra generated by the $\sxx$ Lie
algebra together with $c_-$ and $e_{-1}$, as well as all their 
nonvanishing multiple commutators. The ``classical'' configuration 
space of M-Theory should then be identified with the coset space
\bea 
\tMC = \frac{E_{10}}{\isx} \label{coset3}
\eea

Of course, we will have to worry about the fate of these
symmetries in the quantum theory. Indeed, some quantum version
of the symmetry groups appearing in (\ref{coset3}) must be realized 
on the Hilbert space of third quantized $N=16$ supergravity, such that 
$E_{10}$ becomes a kind of spectrum generating (rigid) symmetry on
the physical states, while the gauge group $\isx$ gives rise to 
the constraints defining them. Because ``third quantization'' here
is analogous to the transition from first quantized string theory
to string field theory, the latter would have to be interpreted 
as multi-string states in some sense (cf. \cite{Witten3} for 
earlier suggestions in this direction; note also that the coset space 
(\ref{coset3}) is essentially generated by half of $E_{10}$, so 
there would be no ``anti-string states''). According to \cite{duality},
the continuous duality symmetries are broken to certain discrete 
subgroups over the integers in the quantum theory. Consequently, 
the quantum configuration space would be the left coset
$$
\tFC = {E_{10}({\bf Z})}\backslash \tMC 
$$
and the relevant partition functions would have to be new kinds 
of modular forms defined on $\tFC$. However, despite recent 
advances \cite{Bakas, Sen}, the precise significance of the
(discrete) ``string Geroch group'' remains a mystery, and it is far from
obvious how to extend the known results and conjectures for finite 
dimensional duality symmetries to the infinite dimensional case
(these statements apply even more to possible discrete hyperbolic
extensions; see, however, \cite{Mizo,GM}). Moreover, recent work 
\cite{KS} confirms the possible relevance of quantum groups 
in this context (in the form of ``Yangian doubles''). 

Returning to our opening theme, more should be said about the
1+10 split, which would lift up the $\LWp\times\isx$ symmetry,
and the ``bein'' which would realize the exceptional geometry 
alluded to in the introduction, and on which $\isx$ would act as 
a generalized tangent space symmetry. However, as long as the 2+9 split 
has not been shown to work, and a manageable realization is not 
known for either $E_{10}$ or $\isx$, we must leave the elaboration 
of these ideas to the future. It could well prove worth the effort.\\ 

\noindent
{\bf Acknowledgments:} The results described in section 2 are based 
on work done in collaboration with S.~Melosch. I would also like 
to thank C.~Daboul, R.W.~Gebert, H.~Samtleben and P.~Slodowy for 
stimulating discussions and comments.

\end{document}